\documentclass[%
 reprint,
 amsmath,amssymb,
 aps,
 prl
]{revtex4-2}

\usepackage[T1]{fontenc}
\usepackage[latin9]{inputenc}
\usepackage{ulem}

\usepackage{graphicx}
\usepackage{epstopdf}
\usepackage{dcolumn}
\usepackage{bm}
\usepackage{hyperref}
\hypersetup{colorlinks=true,citecolor=blue,linkcolor=blue}
\usepackage[caption=false]{subfig}
\usepackage{extarrows}
\usepackage{bbm}
\usepackage{wasysym}

\newcommand{\nicefrac}[2]{\left.#1\middle/#2\right.}

 \global\long\def\Z{\mathbb{Z}}

 \global\long\def\bra#1{\left\langle #1\right|}
 \global\long\def\ket#1{\left|#1\right\rangle }

 \global\long\def\hamil{\mathcal{H}}

 \global\long\def\broket#1#2#3{\bra{#1}#2\ket{#3}}

 \global\long\def\d{\delta}
 
 \global\long\def\L{\mathcal{L}}

 \global\long\def\e{\varepsilon}

\begin{document}

\title{Robustness of the Floquet many-body localized phase in the presence of a smooth and a non-smooth drive}

\author{Asaf A. Diringer$^{1}$ and Tobias Gulden$^{1,2}$} 

\affiliation{$^{1}$ Department of Physics, Technion, 3200003, Haifa, Israel~~\\
 $^{2}$ IST Austria, am Campus 1, 3400 Klosterneuburg, Austria}

\date{\today}

\begin{abstract}
In this work, we investigate how the critical driving amplitude at
the Floquet MBL-to-ergodic phase transition differs between smooth
and non-smooth driving over a wide range of driving frequencies. To this end,
we study numerically a disordered spin-1/2 chain which is periodically
driven by a sine or a square-wave drive, respectively. In both cases, the critical driving amplitude increases
monotonically with the frequency, and at large frequencies, it is
identical for the two drives in the appropriate normalization.
However, at low and intermediate frequencies the critical amplitude
of the square-wave drive depends strongly on the frequency, while the
one of the cosine drive is almost constant in a wide frequency range.
By analyzing the density of drive-induced resonance in a Fourier
space perspective, we conclude that this difference is due to resonances
induced by the higher harmonics which are present (absent) in the
Fourier spectrum of the square-wave (sine) drive. Furthermore, we suggest
a numerically efficient method to estimate the frequency
dependence of the critical driving amplitudes for different
drives, based on measuring the density of drive-induced
resonances. 
\end{abstract}

\maketitle

\section{Introduction}

Remarkable progress in the understanding of closed quantum systems away from equilibrium has taken place in recent years. Specifically, big focus has been put on periodically-driven Floquet systems which provide a new way to engineer known phases as well as unique exotic phases with no equilibrium counterpart \cite{Thouless,Moessner2017,Else2016a,Zhang2017,Kitagawa2010,Lindner2011,Delplace2013,Titum2016,Lohse2016,Nakajima2016,Basov2017,Else2016,rudner2020band,harper2020topology}. Unfortunately, the tendency of a generic driven, interacting system to heat up to a featureless infinite-temperature state poses a fundamental obstacle to Floquet engineering in many-body physics. This is a consequence of the eigenstate thermalization hypothesis (ETH) \cite{SrednickiETH}. While prethermal, quasi-steady states offer a way to transiently stabilize such Floquet-engineered states \cite{Lazarides2014,Abanin2015,DAlessio2014,LindnerBergRudner,Gulden2019,Bukov2016,Berges2004,Eckstein2009,Moeckel2010,Mathey2010,Bukov2015a,Canovi2016,Kuwahara2016,Mori2016,Weidinger,Else2017,PhysRevX.7.011026,PhysRevX.7.011026,PhysRevLett.119.010602}, they lose all memory of initial conditions in the long-time limit. On the contrary, stable steady states can be found in many-body localized systems \cite{Else2016,rudner2020band,harper2020topology,Basko2006,Oganesyan2007,Rehn2016,Lazarides2014}.

In the presence of a sufficiently strong disorder, the emergence of local conserved quantities may induce the many-body localized phase (MBL). These conserved quantities, also known as local integrals of motion or l-bits, stabilize the quantum system with respect to local perturbations. The existence of an extensive set of l-bits leads to a lack of transport and results in the long-term memory of the initial conditions which characterizes the MBL phase \cite{Huse2014,Imbrie2016b,Rademaker2016,DAlessio2016,Nandkishore2015,Serbyn2015}. Studying disordered systems in the presence of a time-periodic drive, the Floquet-MBL phase was discovered: a driven steady-state in which the many-body localized phase survives at all times \cite{PhysRevLett.116.250401,Lazarides2014a}. This was first shown numerically
\cite{Lazarides2015,Ponte2015,Ponte2015a,Zhang2016,Soonwon2017,PhysRevB.96.020201},
and later supported by analytical work \cite{Abanin2016,Burin2017,Gopalakrishnan2016a} and  observed experimentally \cite{Bordia2017,zhang2017observation,choi2017observation,PhysRevX.10.021044}.

However, most numerical papers discuss binary drives, i.e. systems where two static Hamiltonians are exchanged in a non-smooth manner. On the other side, 
typically experiments are performed with smooth drives, and a lot of analytic results were obtained by considering smooth drives \cite{Abanin2016,Burin2017,Gopalakrishnan2016a,Abanin2015,PhysRevB.96.020201,PhysRevB.93.235151}. To bridge this gap between numerics and experiments, in this work we compare the effects of a smooth and a non-smooth drive. Specifically, we study a disordered spin-1/2 XXZ chain in the presence of a sine and a square-wave (sw) drive. We numerically find the critical driving amplitude as a function of driving frequency and analyze the differences in the two phase diagrams.

At large frequencies, the critical amplitude for both drives is virtually identical. However, at intermediate frequencies, there is a large window in which the critical driving amplitude of the sine drive remains almost constant, while the critical amplitude of the square-wave drive decreases as the frequency is decreased, and is clearly smaller than the value for the sine drive. At very small frequencies of the sine drive, as predicted by \cite{Abanin2016,Gopalakrishnan2016a}, we observe a regime where emergent adiabatic Landau-Zener transitions become the main contributor to delocalization. Such transitions are not possible if the system is driven by a square-wave drive, due to the abrupt and non-gradual change of the driving potential. Therefore this qualitatively different regime is absent for the square-wave drive.

In the intermediate- and large-frequency regime delocalization is dominated by drive-induced resonances. The drive hybridizes pairs of eigenstates of the static Hamiltonian. Above a critical density of such resonances, the system delocalizes and the eigenstates of the Floquet Hamiltonian become ergodic. In this paper, we suggest a method of counting drive-induced resonances and estimating the frequency-dependence of the critical amplitude. This method requires only the diagonalization of static Hamiltonians, hence it is far less demanding on the computational resources than direct calculation of the critical amplitude. Furthermore, previously calculated data can be reused to compare the effects of different drives and frequencies.

Contrary to the static case, resonances in time-dependent perturbation theory also include pairs of eigenstates whose energies differ by amounts close to integer multiples of the driving frequency, $\Delta E\approx k\omega$ with $k\in\Z$. There are two main processes that may cause such a resonance: either absorption/emission of a single photon with energy $k\omega$, or $k$ photons with energy $\omega$ each (or any combination of photons). However, the former process only appears if the $k$-th Fourier component of the drive is non-zero. Here lies the main difference between the sine and the square-wave drive: the Fourier series of the smooth sine drive only has one component with the fundamental frequency $\omega$, while the Fourier series of the non-smooth square-wave drive contains all (odd) multiples of $\omega$. Hence, only the square-wave drive can couple pairs of eigenstates with $\Delta E\approx k\omega$, $|k|>1$, at first order in perturbation theory. The sine drive only couples such pairs through higher-order processes whose amplitude is exponentially suppressed in the energy difference.

Throughout this paper, we normalize the driving strength so that the first Fourier component of both drives is identical. Hence, as compared to the sine drive, the square-wave drive has additional terms which are the higher Fourier components. Therefore the square-wave drive creates additional resonances, and we expect it to delocalize the system further compared to the sine drive of the same amplitude and frequency. I.e., the critical driving amplitudes $g^{c}$ should obey the relation
\begin{equation}
    g_{sw}^{c} (\omega) \leq g_{sin}^{c}(\omega).
\end{equation}

Due to the extensive set of l-bits, it is useful to think of the MBL phase as composed of a set of local subsystems of finite spatial extent which have a discrete spectrum of bounded bandwidth $\kappa W$ (where $W$ is the disorder strength and $\kappa$ is a dimensionless constant which depends on microscopic details of the system) \cite{Abanin2016, Lazarides2015}. In the presence of a drive which is composed of local terms, the local subsystems are only coupled to neighboring subsystems. If the system is driven at frequencies larger than the typical energy width of the coupled blocks, the additional resonances induced by the time-dependent drive are scarce. This leads to the stability of the MBL phase at large $\omega$   \cite{Abanin2016,Lazarides2015}.

Conversely, two eigenstates $\ket m,\ket n$ with substantial energy difference $\Delta E_{m,n}$ differ by many l-bits. Thus the matrix element $\broket m{\hat{V}}n$ connecting the two states with the local operator $\hat{V}$ is small and decays exponentially with the number of different l-bits. Hence, if the driving frequency approaches the local bandwidth resonances induced by the higher Fourier components with $|k\omega|>\kappa W$ are suppressed. Therefore the role of the higher Fourier components of the square-wave drive is negligible and we expect a similar value of the critical driving amplitude for both the sine and square-wave drive:
\begin{equation}
    g_{sw}^{c}\left(\omega\right)\approx g_{sin}^{c}\left(\omega\right)
    \text{ for }\omega\apprge\kappa W.\label{eq:definition kappa}
\end{equation}

\begin{figure}[t]
    \includegraphics[width=1\columnwidth]{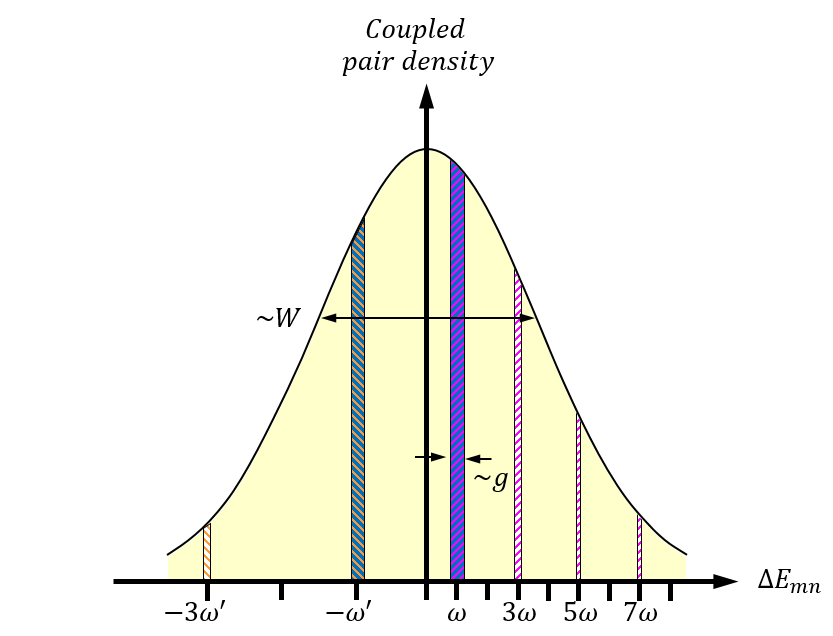}
    \caption{Schematic drawing of the density of pairs of eigenstates which are coupled by the drive, as a function of their energy difference $\Delta E_{mn}$. Pairs of states with $\Delta E_{mn}\approx k\omega$  become resonant if the corresponding Fourier component of the drive $f_k$ is non-zero, cf. Eq. \eqref{eq:resonance}. At small driving frequency $\omega$ (right side) the square-wave drive (red) couples many more pairs resonantly than the sine drive (blue). This means, if $\omega$ is small compared to $W$ (the width of the Gaussian density of pairs), there is a large difference in the density of resonances caused by the two drives. However, for a large driving frequency $\omega'$ (left side) there is only a small difference between the two drives because there are only few pairs of states with energy difference $\Delta E_{mn}\approx \pm3\omega'$ (green/orange).}
    \label{fig:schematic}
\end{figure}

On the other hand, at frequencies much smaller than the local bandwidth, some of the higher Fourier components of the square-wave lie well within the local bandwidth, $|k\omega|\leq\kappa W$ for sufficiently small $\omega$, and induce additional resonances. This is visualized in Fig. \ref{fig:schematic}. Therefore we expect a more drastic reduction of the critical driving amplitude of the square-wave drive as compared to the sine drive when the driving frequency is small, leading to 
\begin{equation}
    g_{sw}^{c}\left(\omega\right)<g_{sin}^{c}\left(\omega\right)\text{ for }\omega<\kappa W.
\end{equation}

In the following, these predictions are tested numerically by applying two different drives to a commonly used model in the study of the MBL phase transition.

\section{Model}

We consider the spin-$\nicefrac{1}{2}$ XXZ chain with periodic boundary conditions in a disordered longitudinal field. The static Hamiltonian is 
\begin{align}
\hamil_{0} & =\sum_{i}\e_{i}\sigma_{i}^{z}+U\sum_{i}\sigma_{i}^{z}\sigma_{i+1}^{z}\nonumber \\
 & +\frac{J}{2}\sum_{i}\left(\sigma_{i}^{+}\sigma_{i+1}^{-}+H.c.\right)\label{eq:static Hamiltonian}
\end{align}
where $U$ is the nearest-neighbor interaction strength, $J$ the spin interaction, and $\e_{i}$ the random on-site potential which is drawn uniformly and independently from the interval $\left[-W,W\right]$. Via a Jordan-Wigner transformation, this model is related to interacting systems of both, fermions or hard-core bosons. Such models were studied extensively and feature a transition to a many-body localized phase for sufficiently strong disorder \cite{Oganesyan2007,Pal2010,Iyer2013,Kjall2014}. We normalize the energy scale by setting the strength of the nearest-neighbor interaction to be $U=1$, and we set $W=3$ for the disorder strength and $J=0.2$ for the hopping term. In the appendix we present additional data for the case $J=0.6$ which puts the static Hamiltonian close to the MBL-to-ergodic phase transition.

We study the Floquet-MBL transition when the system is subjected to a time-periodic drive with period $T=\nicefrac{2\pi}{\omega}$, 
\begin{align}\label{eq:Tot Hamiltonian}
    \hamil\left(t\right) & =\hamil_{0}+V\left(t\right),\\
    V\left(t\right) & =V\left(t+T\right).\nonumber 
\end{align}
The drive is implemented by a translationally invariant sum of local operators, 
\begin{align}
    V\left(t\right) & =g\cdot f\left(t\right)\sum_{i}\left(\sigma_{i}^{+}\sigma_{i+1}^{-}+H.c.\right)\nonumber \\
    & +g\cdot f\left(t\right)\sum_{i}\left(-1\right)^{i}\sigma_{i}^{z}\label{eq:general drive}
\end{align}
where $f(t)=f(t+T)$ is a periodic function of time and $g$ is the driving amplitude.

As mentioned above, we choose one smooth and one non-smooth driving function $f(t)$, and analyze the different behavior of the critical driving amplitude $g^{c}$. As the most basic example, for a smooth periodic drive we choose a sine function, 
\begin{align}
    f_{sin}\left(t\right) & =\sin\left(\omega t\right)\nonumber \\
    & =\frac{1}{2i}\left(\exp\left(i\omega t\right)-\exp\left(-i\omega t\right)\right).\label{eq:smooth drive}
\end{align}
For a non-smooth and discontinues drive, we use the periodic square-wave or rectangular function which gives the binary drive used in previous numerical studies of the Floquet-MBL to ergodic phase transition \cite{Lazarides2015,Ponte2015,Ponte2015a,Zhang2016,Soonwon2017,PhysRevB.96.020201}, 
\begin{align}
    f_{sw}\left(t\right) & =\text{\ensuremath{\frac{\pi}{4}}sgn}\left[\sin\left(\omega t\right)\right] \\
    & =\sum_{k\in\Z^{\text{odd}}} \frac{1}{2ik}
    \exp\left(ik\omega t\right),
    \label{eq:sw drive}
\end{align}
where the sum is taken over all odd integers $k$. 

The most important difference between these two drives lies in their representation as a Fourier series. While the sine function has only a single Fourier component, the square-wave function has infinitely many Fourier components which decay proportional to $\nicefrac{1}{k}$. Note that we normalize $f_{sw}$ so that its first Fourier component is equal to the corresponding Fourier component of a sine drive of the same driving amplitude $g$.

\begin{figure}[t]
    \includegraphics[width=1\columnwidth]{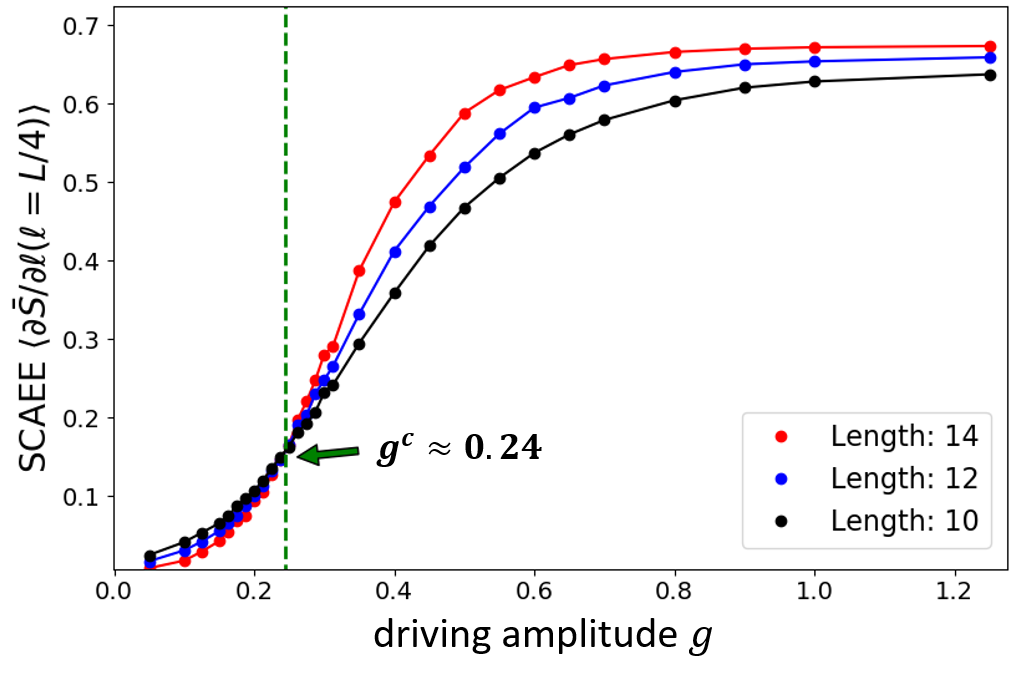}
    \caption{The SCAEE as a function of driving amplitude, for a sine drive at frequency $\omega=1$. We average the CAEE over 1500, 800, 400 disorder realizations (2400, 1200, 500 at amplitudes in the vicinity of $g^{c}$) for chains of length $L=10,12,14$, respectively. Further averaging was achieved by randomly drawing $\approx50$ eigenstates of the single-cycle evolution operator at each realization. The crossing of the three lines gives the critical driving amplitude $g^{c}\simeq0.24$.}
    \label{fig:SCAEE}
\end{figure}

\section{Numerical Methods}

In this section, we describe how we calculate the critical driving amplitude and the density of drive-induced resonances.

\subsection{Critical driving amplitude}

\begin{figure*}
    \includegraphics[width=1.8\columnwidth]{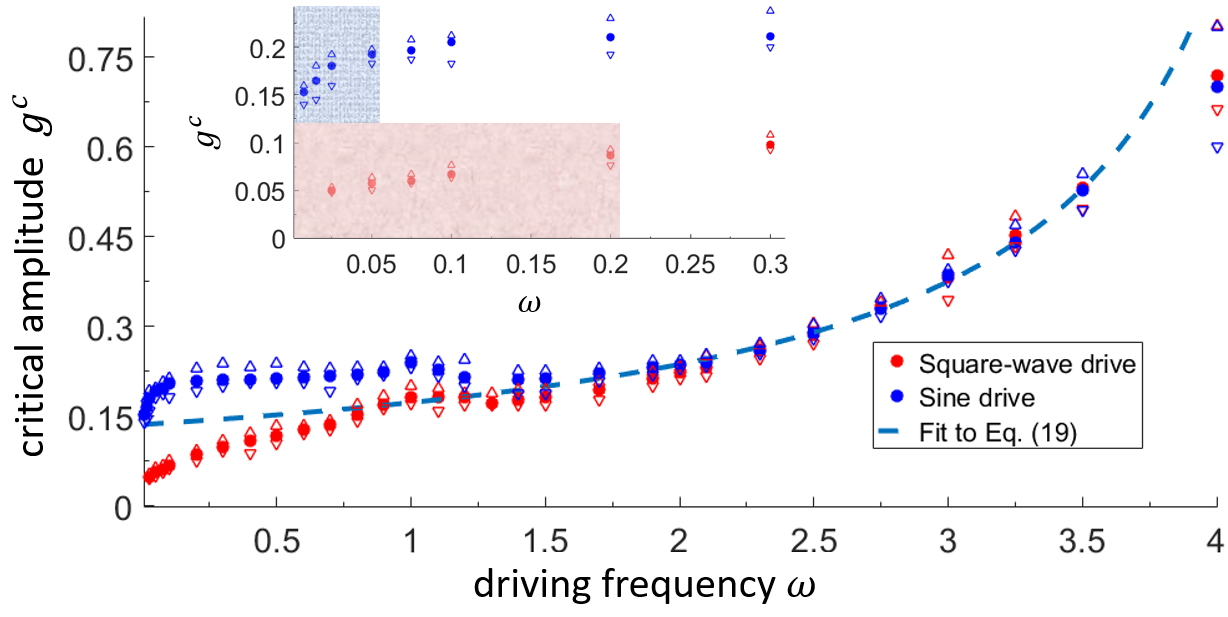}
    \caption{Critical driving amplitudes $g^{c}(\omega)$ of the sine (blue) and square-wave (red) drives as a function of the driving frequency $\omega$. The static Hamiltonian $\mathcal{H}_0$ is deep in the MBL phase $\left(J=0.2,W=3.0\right)$. In both cases the critical amplitude was inferred by finite-size scaling of the SCAEE, cf. Fig. \ref{fig:SCAEE}. The dashed line is the fit to Eq. \eqref{eq:gcfit} for $1.7\leq\omega\leq3.5$ with parameters $\beta=0.21$ and $\nicefrac{n_{res}^{c}}{\alpha}=0.13$.\protect \\
    Inset: A closer look at small frequencies $\omega$. Here the critical amplitude of the sine drive becomes larger than the driving frequency, $g_{sin}^{c}(\omega)>\omega$. This leads to the appearance of adiabatic Landau-Zener transitions which causes a deviation from the almost-constant behavior of the critical driving amplitude. The shaded areas mark where we expect significant finite-size effects for the sine (blue) and square-wave (red) drive, respectively.}
    \label{fig:gcritical}
\end{figure*}

We want to find the critical amplitude $g_{sin,sw}^{c}(\omega)$ at different driving frequencies $\omega$. To this end we calculate the time-evolution operator of the full Hamiltonian (\ref{eq:Tot Hamiltonian}) over a single driving period, $U\left(T\right)=\mathcal{T}e^{-i\int_{0}^{T}H(t)dt}$, where $\mathcal{T}$ is the time-ordering operator. We determine the Floquet eigenstates through exact diagonalization of $U(T)$ \cite{SciPostPhys.2.1.003}. Subsequently, we determine the critical driving amplitude
through finite-size scaling of the slope of the cut-averaged entanglement entropy (SCAEE) \cite{PhysRevB.94.184202}. CAEE is the entanglement entropy as a function of the size $\ell$ of a simply-connected subsystem, averaged over all possible locations inside the system. The CAEE is further averaged over $\approx50$ randomly chosen Floquet eigenstates from each disorder realization. SCAEE is the slope of the CAEE at $\ell=\nicefrac{L}{4}$,
\begin{equation}
    \frac{\partial\bar{S}\left(\ell\right)}{\partial\ell}|_{\ell=\nicefrac{L}{4}}\equiv\left(\frac{1}{L}\frac{\partial}{\partial\ell}\sum_{x=1}^{L}S\left(x,\ell\right)\right){}_{\ell=\nicefrac{L}{4}},
    \label{eq:SCAEE}
\end{equation}
which we evaluate with a quadratic spline fit. Here $S\left(x,\ell\right)$ is the averaged entanglement entropy of the simply connected subsystem of size $\ell$ centered around site $x$. The approximate critical driving strength is found through finite-size scaling. In a fully localized system, the SCAEE goes to zero in the thermodynamic limit, while in a delocalized system the slope of the entanglement entropy tends to $\ln2$. Hence a decrease of the SCAEE as system size is increased means the system is in the localized phase, while an increase and an approach towards the value of $\ln2$ implies that the system is ergodic. Therefore, as illustrated in Fig. \ref{fig:SCAEE}, the crossing between curves of different system size marks the critical driving strength \cite{PhysRevB.94.184202}. A summary of our results for the critical amplitude of both drives is shown in Fig. \ref{fig:gcritical}.

\subsection{Density of resonances}\label{resonances_Counting}

Additionally we calculate the density of first-order resonances induced by the sine and square-wave drives at different driving frequencies $\omega$. The drives in Eq. (\ref{eq:general drive}) are characterized by driving amplitude $g$ and frequency $\omega$, shape of the driving function $f(t)$, and the operator 
\begin{equation}
    \hat{V}=\sum_{i}\left(-1\right)^{i}\sigma_{i}^{z}+\left(\sigma_{i}^{+}\sigma_{i+1}^{-}+H.c.\right).
\end{equation}
The main difference between the sine and square-wave drive appears in their temporal Fourier series, $f\left(t\right)=\sum_{k}f_{k}\exp\left(ik\omega t\right)$, with
\begin{equation}
    \left|f_{k}\right|=\begin{cases}
    \nicefrac{1}{2} & ,\left|k\right|=1\\
    0 & ,\text{otherwise}
    \end{cases} \label{eq:Cosine Fourier}
\end{equation}
for the sine drive in Eq.~\eqref{eq:smooth drive}, and 
\begin{equation}
    \left|f_{k}\right|=\begin{cases}
    \nicefrac{1}{2k} & ,k\text{ is odd}\\
    0 & ,\text{otherwise}
    \end{cases} \label{eq:SW Fourier}
\end{equation}
for the square wave drive in Eq.~\eqref{eq:sw drive}, respectively.

In order to determine the density of first-order resonances, we first determine the eigenstates of the static Hamiltonian $\hamil_{0}$ in Eq. \eqref{eq:static Hamiltonian}. Next, we calculate the matrix element $\broket m{\hat{V}}n$ as well as the energy difference $\Delta E_{mn}=\broket m{\hamil_{0}}m-\broket n{\hamil_{0}}n$ for all pairs of eigenstates $\ket m,\ket n$. We call a pair of eigenstates \textit{resonant at first order} if they are connected by a single application of the driving operator $\hat{V}$ (i.e. their Hamming distance is 1), and they satisfy the inequality
\begin{equation}
    g\cdot\max_{k\in\Z}\left|\frac{f_{k}\broket m{\hat{V}}n}{\Delta E_{mn}-k\omega}\right|>\d,\,k\in\Z.\label{eq:resonance}
\end{equation}
where $\delta$ is an arbitrary number of order 1.

For both drives, we count the number of states which satisfy the inequality \eqref{eq:resonance} as a function of driving amplitude $g$ and frequency $\omega$, where we arbitrarily choose $\delta=\sqrt{3}/2$  and normalize the result by the system size to get the density of resonances. This procedure is then repeated over many disorder realizations. The results are shown in Fig. \ref{fig:resonances}.

Unlike in calculating the SCAEE, we note that the resonance counting scheme avoids the costly calculation of the Floquet evolution operator and only requires diagonalization of the static Hamiltonian $\mathcal{H}_0$ at a single system size. After the static eigenstates are found, the matrix elements $\broket m{\hat{V}}n$ may be calculated for many different operators $\hat{V}$. Furthermore, the critical amplitudes $g^c$ as a function of driving frequency $\omega$ of many drives which differ only by temporal form, are easily estimated with very little demand on computational resources.

\section{Results and Discussion}\label{sec:results}

The main goal of this paper is a comparison of the critical amplitude $g^c$ of the MBL-to-ergodic transition between two different drives, a smooth sine drive and a non-smooth square-wave drive. Fig.~\ref{fig:gcritical} summarizes these results. There we can distinguish between three qualitatively different regimes, which we analyze in the following.

In the large frequency regime, $\omega>\Omega^{*}\approx 2.0$, the critical amplitude of both drives is equal and prominently increases with the driving frequency. On the other hand, as the driving frequency is decreased below the threshold $\omega<\Omega^{*}$, the critical driving amplitudes of the two drives separate. The square-wave drive destabilizes the localized phase at a clearly smaller driving amplitude than its sine counterpart, i.e. $g_{sw}^{c}(\omega)<g_{sin}^{c}(\omega)$. In this regime $g_{sw}^{c}(\omega)$ monotonically decreases with decreasing $\omega$, while $g_{sin}^{c}(\omega)$ shows little to no variation until $\omega\approx0.1$. It was predicted that at such small frequencies adiabatic Landau-Zener (LZ) transitions become a major contributor to delocalization, resulting in a reduced critical driving amplitude \cite{Abanin2016,Gopalakrishnan2016a}.

\subsection{Critical density of resonances and delocalization}
\begin{figure*}[t]
    \includegraphics[width=1.8\columnwidth]{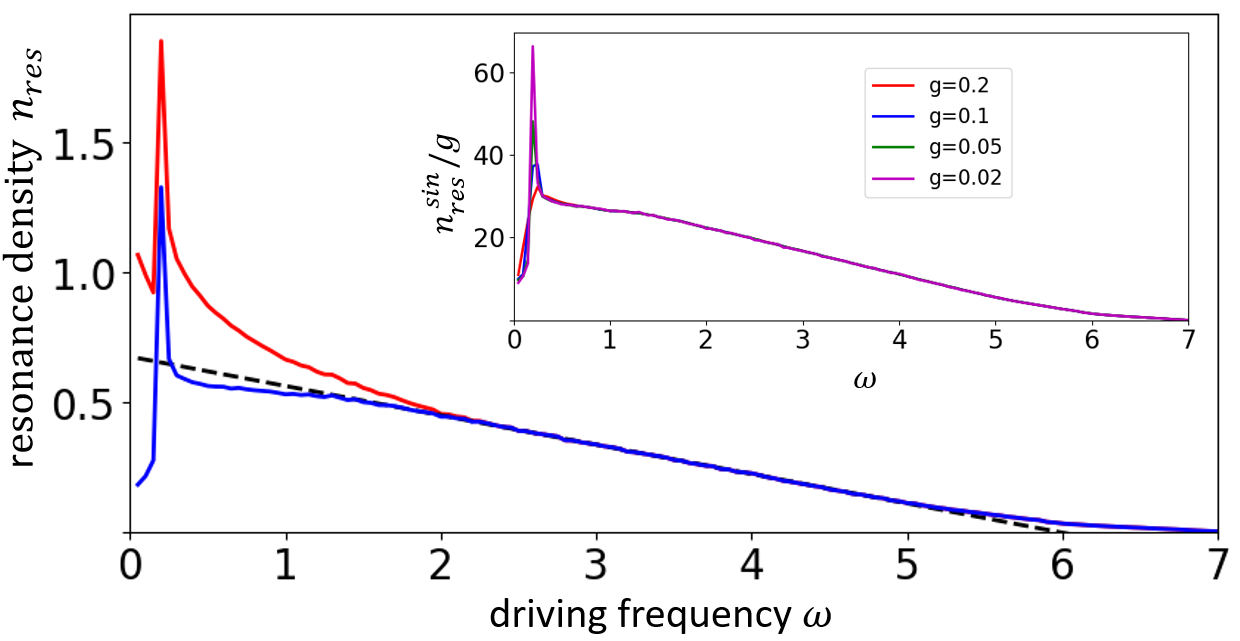} 
    \caption{The density of drive-induced resonances per unit length, as defined in Eq. (\ref{eq:resonance}). The system is driven by a sine (blue) or square-wave (red) drive with amplitude $g=0.2$, and we choose $\delta=\nicefrac{\sqrt{3}}{2}$. The system parameters are $J=0.2,\,W=3.0$. The data was obtained by counting the number of eigenstate pairs of a spin chain of length $L=10$ which satisfy the inequality \eqref{eq:resonance}, averaged over $50,000$ disorder realizations. The black dashed line is a linear fit in the range $1.7\leq\omega\leq5$ of $n_{res}^{sin}(\omega)$, according to Eq. (\ref{eq:nresfit}). The extracted parameters are $\beta=0.17,\,\alpha=0.075$.\\
    Inset: The density of resonances induced by the sine drive for different driving amplitudes $g$. Normalizing the density of resonances with the driving amplitude leads to a collapse onto a single curve. This shows that the density of resonances is proportional to the driving strength, cf. Eq. \eqref{eq:nresfit}. The peak at $\omega\approx0.2$ does not collapse, i.e. it is not proportional to the driving amplitude. This indicates that it does not originate from drive-induced resonances, but from the static Hamiltonian.}
    \label{fig:resonances}
\end{figure*}
To explain the results in Fig. \ref{fig:gcritical}, we propose a model of delocalization in a Floquet system which mainly depends on reaching a critical, frequency-independent density of resonances, $n_{res}^{c}$. This is based on a similar model for the MBL-to-ergodic transition in time-independent systems proposed in \cite{Imbrie2016,Imbrie2016b,PhysRevLett.121.140601}.

In a weakly driven system with $g\ll\omega,W$ a pair of eigenstates $\ket m,\ket n$ can only hybridize if their energy difference is close to an integer multiple of the
driving frequency, 
\begin{align}
    \Delta E_{mn} & =\broket m{\hamil_{0}}m-\broket n{\hamil_{0}}n.\nonumber \\
     & \approx k\omega,\,k\in\Z
\end{align}
Furthermore the matrix element between these two states must be non-zero, $\broket m{\hat{V}}n\neq0$, as well as the corresponding Fourier component of the drive, $f_k\neq0$. This is reflected in the definition of a resonant pair in  Eq.~\eqref{eq:resonance}.

An important characteristic of the localized phase is that the matrix element of a local perturbation between a pair of static eigenstates typically decays exponentially with the number of l-bits by which the two states differ \cite{Rademaker2016,Serbyn2013,Ros2014}. Furthermore, as shown in \cite{PhysRevB.90.224203,Gopalakrishnan2015} most l-bits of a system which is deep in the MBL phase are well approximated by a few or even by a single physical bit (p-bit). This holds for the strongly localized Hamiltonian $\hamil_{0}$ in Eq. \eqref{eq:static Hamiltonian}.

The drive defined in Eq. (\ref{eq:general drive}) only allows for a single spin exchange between two neighboring p-bits. Hence, at first order in the driving amplitude, it only leads to strong coupling between pairs of static eigenstates $\ket m,\ket n$ which are mostly identical except for the two corresponding l-bits. The energy difference $\Delta E_{mn}$ of such pairs of eigenstates is dominated by the difference in on-site potential of the corresponding p-bits. As this energy is proportional to the disorder strength $W$, the effective amplitude for a rearrangement of $n$ l-bits goes (up to a combinatorial factor) as $\tilde{g_{n}}\sim\nicefrac{g^{n}}{W^{n-1}}$\cite{Gopalakrishnan2016a}. 

Therefore, for a system with $g\ll W$, such higher-order processes are exponentially suppressed and drive-induced heating is expected to be dominated by first-order processes. Fig. \ref{fig:resonances} shows the density of first-order resonances induced by the two drives, estimated by our resonances counting scheme in Eq. \eqref{eq:resonance}.

Inspired by the discussion of static systems in \cite{PhysRevLett.121.140601,Imbrie2016b} we assume that delocalization occurs when the density of resonances induced by the drive surpasses a critical value which is independent of the driving frequency. The resonance criterion in Eq. \eqref{eq:resonance} and the inset in Fig. \ref{fig:resonances} show that the density of resonances is proportional to the driving amplitude,
\begin{equation}
    n_{res}\left(g,\omega\right)=g\cdot n_{res}\left(g=1,\omega\right).\label{eq:proportionaltog}
\end{equation}
Combined with the hypothesis of a critical, drive-independent density of resonances $n_{res}^{c}$ we can express the critical driving amplitude as 
\begin{equation}
    g^{c}(\omega)=\frac{n_{res}^{c}}{n_{res}(\omega)}\sim n_{res}(\omega)^{-1}.\label{eq:gcomega}
\end{equation}
Hence, the frequency dependence of the density of resonances translates directly to the critical driving amplitude.

In Fig. \ref{fig:resonances} we find that for $1.7\apprle\omega\apprle5.0$ the density of resonances induced by the sine drive decreases linearly with the driving frequency, 
\begin{equation}
    n_{res}^{sin}(\omega,g)\approx\alpha g(1-\beta\omega).\label{eq:nresfit}
\end{equation}
This linear behavior is likely due to our implementation of the disorder by a uniformly distributed random variable. Applying Eq. (\ref{eq:nresfit}) to (\ref{eq:gcomega}) gives 
\begin{equation}
    g_{sin}^{c}(\omega\apprge1.7)=\frac{\nicefrac{n_{res}^{c}}{\alpha}}{1-\beta\omega}.\label{eq:gcfit}
\end{equation}
In Fig. \ref{fig:gcritical} we fit the numerical data to Eq. (\ref{eq:gcfit}), and find good agreement with the data points within the range $1.7\apprle\omega\apprle3.5$. We restrict ourselves to this upper bound as an analysis of the many-body bandwidth indicates that finite-size effects become more substantial at larger driving frequencies (cf. e.g. Supplement to \cite{Lazarides2015}). From this fit we extract the fit parameter $\beta=0.21$, while the linear fit in Fig. \ref{fig:resonances} yields $\beta=0.17$. This quantitative discrepancy can be explained by the fact that our scheme of counting resonances only takes into account first-order resonances, while especially at larger driving amplitudes higher-order resonances also contribute to delocalization. Nevertheless, the good qualitative agreement is an indication that delocalization is indeed controlled by exceeding a critical density of resonances.

\subsection{Frequency-dependence of the critical driving amplitude}

In Fig. \ref{fig:gcritical} we note that for $\omega>\Omega^{*}\approx2.0$ the critical amplitudes of the two drives agree, $g_{sin}^{c}=g_{sw}^{c}$. We relate this to the fact that the higher Fourier components of the square-wave drive can not induce additional resonances. Specifically, for a square-wave drive once $3\omega\apprge\kappa W$ (with $\kappa$ as defined in Eq.~(\ref{eq:definition kappa})) the density of induced resonances is strongly dominated by its first Fourier component. Thus, driving at frequencies greater than this value, the density of resonances induced by these two drives is the same, $n_{res}^{sw}(\omega)=n_{res}^{sin}(\omega)$. This is reflected in Fig. \ref{fig:resonances}. Combined with our hypothesis that delocalization happens at a critical, drive-independent density of resonances, we obtain  $g_{sin}^{c}\left(\omega\right)\approx g_{sw}^{c}\left(\omega\right)$ in this frequency regime (as is observed in Fig. \ref{fig:gcritical}).

On the other hand, at driving frequencies $0.2\apprle\omega\apprle\Omega^{*}$ the two drives cause very different behavior. In Fig. \ref{fig:resonances} the density of resonances induced by the square-wave drive increases drastically as $\omega$ is decreased, agreeing with the decrease of $g_{sw}^{c}$ at small $\omega$ in Fig. \ref{fig:gcritical}. On the contrary, the density of resonances induced by the sine drive is only weakly dependent on the frequency and seems to saturate, corresponding to a region of weak frequency-dependence of $g_{sin}^{c}$ around $0.2\apprle\omega\apprle\Omega^{*}$ found in Fig. \ref{fig:gcritical}. In this frequency range, the higher Fourier components of the square-wave drive can also couple pairs of eigenstates with energy difference $\Delta E_{mn}\approx k\omega$, $|k|>1$, which are not coupled at first order by the sine drive \footnote{Note that resonances between such pairs of eigenstates can occur for the sine drive as well, however they appear only through higher-order processes in the driving amplitude. Therefore, while the effective amplitude for such resonant transitions induced by the square-wave drive is proportional to the corresponding Fourier component which decays as $1/\Delta E_{mn}$, the effective amplitude of the sine drive for the same transitions decays exponentially with the energy difference.}. This is visualized in Fig. \ref{fig:schematic}.
Hence the total density of first-order resonances induced by the square-wave drive exceeds that of the sine drive,  $n_{res}^{sw}(\omega)>n_{res}^{sin}(\omega)$ for $\omega<\Omega^{*}$. Consequently this leads to a smaller critical driving amplitude of the square-wave drive, $g_{sw}^{c}(\omega)<g_{sin}^{c}(\omega)$ (cf. Fig. \ref{fig:gcritical}).

At very small driving frequencies $\omega\apprle0.2$ we observe that the critical amplitude of the sine drive exceeds the driving frequency, $g_{sin}^{c}(\omega)>\omega$. This regime was also discussed in Ref.~\cite{Abanin2016}. \textit{Ibid.} it was argued that $g\ll\omega$ is a necessary condition for localization for a generic drive, but for a smooth drive at very low frequencies localization may persist for $g>\omega$ (under the condition $g\ll W$). The reason is that for such a slowly-varying potential $V(t)$ delocalization is additionally caused by the appearance of adiabatic Landau-Zener (ALZ) transitions, and not only resonant coupling of eigenstates. In this regime, if the driving frequency is lowered further, the density of ALZ transitions increases. This leads to a reduction of $g_{sin}^{c}(\omega)$ as compared to the intermediate frequency regime where $g_{sin}^c$ is almost constant as function of $\omega$. In Fig.~\ref{fig:gcritical} we observe this downturn of the critical driving amplitude. However, we want to note that finite-size effects play a more significant role in this low-frequency regime, therefore while the data is an indication for the appearance of ALZ transitions, it is not conclusive.

In contrast, the square-wave drive is constant in time except for discontinuous jumps. Therefore it only causes \textit{diabatic} Landau-Zener transitions (instead of ALZ) which don't contribute to heating and delocalization \cite{Abanin2016}. However, at very small driving frequencies even more Fourier modes cause resonant coupling of eigenstates. Therefore this regime is not qualitatively different from the intermediate frequency regime $0.2\apprle\omega\apprle\Omega^{*}$, and it is expected that the ratio of the critical driving frequency of a square-wave drive over frequency is always smaller than some constant, $\nicefrac{g_{sw}^{c}(\omega)}{\omega}<c$ \cite{Abanin2016}. Accordingly we see a monotonic decrease of $g_{sw}^{c}(\omega)$ in Fig.~\ref{fig:gcritical}. However, as noted above the results in this regime are not conclusive due to finite-size effects.

Finally we must reconcile the sharp peak of $n_{res}$ at $\omega\approx0.2$ for both drives. While such a peak should correspond to a sudden drop of $g^{c}(\omega\approx0.2)$, such peculiar behavior is absent from Fig. \ref{fig:gcritical}. The reason is that the peak originates from the interaction term in the static Hamiltonian, cf. Eq. \eqref{eq:static Hamiltonian}, and not from the drive. For all simulations, we chose $J=0.2$ for the interaction strength. This causes the hybridization of states with energy difference $\Delta E_{mn}\approx0.2$. In our counting scheme these are indistinguishable from resonances induced by the drive, which causes the apparent large density of resonances at $\omega\approx0.2$. From the inset in Fig. \ref{fig:resonances} we can see that this peak is independent of the driving strength $g$ which confirms that it is not caused by the drive but by the static Hamiltonian.

\section{Summary}

In this paper, we presented a numerical study of a Floquet-MBL system subjected to a sine and a square-wave drive and analyzed the different effects on the critical driving amplitude and the density of induced resonances. We use these two common examples in this work, however the results also apply to different smooth or non-smooth drives. With this, we bridge the gap between previous numerical work which uses non-smooth drives (square-wave or periodic kicking) and analytic and experimental results which use smooth drives.

We compared the stability of the MBL phase under the influence of these two driving protocols. Fig. \ref{fig:gcritical} shows our results for the critical driving amplitude for a smooth sine and a non-smooth square-wave drive. Remarkably, at large frequencies $\omega>\Omega^{*}$ the shape of the drive does not influence the delocalization transition. However, at lower frequencies, the square-wave drive delocalizes the system at much smaller driving amplitude than the sine drive. This can be attributed to the effect of higher harmonics in the drive which are only present in the square-wave drive. While the critical amplitude of the square-wave drive decreases significantly when lowering the driving frequency, the critical amplitude of the sine drive is almost constant in a wide range of frequencies.

This behavior can be explained by a model of a critical density of resonances at the MBL-to-ergodic transition. The $k$-th Fourier component of the drive can only couple a pair of states $\ket m,\ket n$ resonantly if their energy difference is $\Delta E_{m,n}\approx k\omega$. However, at large frequencies, this energy is easily bigger than the local bandwidth for the higher Fourier components $k\geq2$. Hence these do not induce additional resonances, so in this regime the sine drive and the square-wave drive are indistinguishable. At small frequencies, the higher Fourier components of the square-wave drive cause additional resonances which lead to a lower critical amplitude as compared to the sine drive.

Furthermore, we showed that studying the density of resonances provides a good estimate for the frequency-dependence of the critical driving amplitude. This is in agreement with the natural assumption that delocalization happens when the drive-induced resonances reach a critical density, and with previous work on static MBL systems~\cite{Imbrie2016,Imbrie2016b,PhysRevLett.122.040601,PhysRevLett.121.140601}.
Among the advantages of this scheme is the fact that calculating the density of resonances is much less demanding on the computational resources than exact diagonalization of the full Floquet operator. In fact, for a comparison of different driving schemes, it is sufficient to calculate the eigenstates of the static Hamiltonian once. These can be subsequently used to calculate the density of induced resonances for all drives of interest, with different frequencies and shapes. This allows for easy comparison of the ability of different drives to destabilize the MBL phase, and Eq. \eqref{eq:gcomega} gives a qualitative prediction for the frequency-dependence of the critical driving amplitude. In this work, we compared drives with different temporal shapes. It would be interesting to see whether this comparison also gives a good prediction for the relative driving strength if the drive operator $\hat{V}$ was changed.

\section*{Acknowledgements}

We thank Yevgeny Bar Lev, Eyal Bairey and Barak Katzir for illuminating discussions and their many insights.  Especially, the authors thank Netanel Lindner for his support and many comments throughout this project. We are further grateful to Maksym Serbyn for reading the manuscript and providing good feedback and suggestions. We acknowledge financial support from the Defense Advanced Research Projects Agency through the DRINQS program, grant No. D18AC00025.  
TG was in part supported by
an Aly Kaufman Fellowship at the Technion. TG acknowledges funding from the Institute of Science and Technology (IST) Austria, and from the European Union's Horizon 2020 research and innovation program under the Marie Sk\l{}odowska-Curie Grant Agreement No. 754411.

\appendix

\section{Appendix- Static system near the MBL transition\label{large J}}

\begin{figure}
    \subfloat[ \label{fig:Not-deep-critical amplitude}]{\includegraphics[width=0.98\columnwidth]{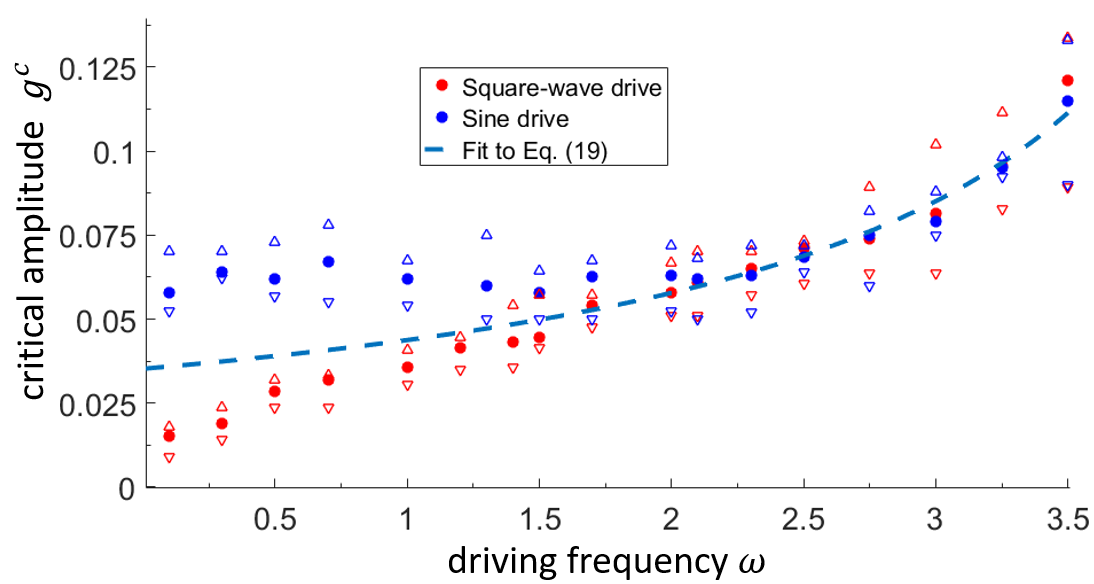}
    }\\
    \subfloat[\label{fig:Not-deep-resonances}]{\includegraphics[width=0.98\columnwidth]{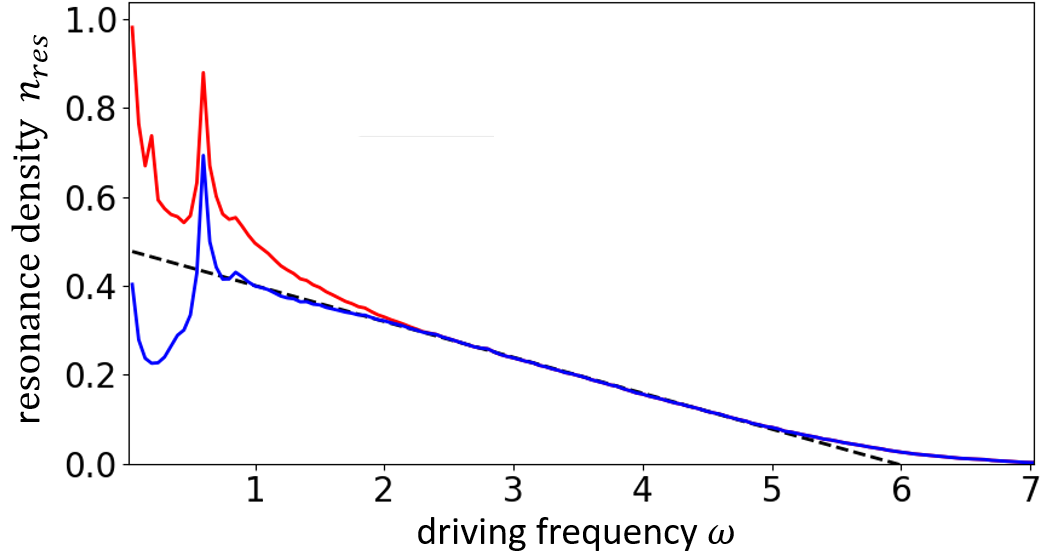}
    }
    \caption{Results when the static Hamiltonian $\mathcal{H}_0$ is near the MBL-to-ergodic transition ($W=3.0,\,J=0.6$).\\
    (a) Critical driving amplitudes $g_{sin/sw}^{c}(\omega)$ of the sine (blue) and square-wave (red) drives as a function of the frequency inferred by SCAEE. The dashed line is the fit to Eq. \eqref{eq:gcfit} for $2.0\leq\omega\leq3.5$ with $\beta=0.19$ and $n_{res}^{c}/\alpha=0.035$. \\
    (b) Density of drive-induced resonances per unit length, as defined in Eq. (\ref{eq:resonance}) with $\delta=\nicefrac{\sqrt{3}}{2}$. The system is driven by a sine (blue) or square-wave (red) drive with amplitude $g=0.02$. The data was averaged over $40,000$ disorder realizations of a spin chain of length $L=10$. The black dashed line is a linear fit in the range $1.9\leq\omega\leq5$ of $n_{res}^{sin}(\omega)$, according to Eq. \eqref{eq:nresfit} with $\beta=0.17,\,\alpha=0.048$.}
    \label{fig:Not-deep}
\end{figure}

In the main text we focus on the case $J=0.2$ which puts the static Hamiltonian deep in the localized phase. Here we offer additional results with the choice of parameter $J=0.6$. This sets the static Hamiltonian in the localized phase but close to the MBL-to-ergodic transition. Our results for the critical driving amplitude and density of induced resonances are summarized in Fig. \ref{fig:Not-deep}. 

As the static system is much closer to the phase transition the stability of the localized phase weakens and the critical driving amplitude decreases. Nevertheless, the manner in which the critical amplitude $g^{c}$ and the density of resonances $n_{res}$ depend on the driving frequency is qualitatively the same as our results in Fig. \ref{fig:gcritical}. Most notably, we see the agreement between the sine and the square-wave drive at large frequencies, and the appearance of a critical frequency $\Omega^{*}$ below which the curves start to diverge. Furthermore, we see that the peak in the density of resonances has shifted from around $\omega=0.2$ in Fig. \ref{fig:resonances} to around $\omega=0.6$ in Fig. \ref{fig:Not-deep}, in accordance with the change of the interaction parameter from $J=0.2$ to $J=0.6$. This shift increases our confidence that this peak stems rather from resonances in the static Hamiltonian than from drive-induced resonances. Hence it has no relevance for the drive-induced phase transition, and we don't see a feature in the critical driving amplitude at that frequency.

\bibliography{general} 

\end{document}